\pdfoutput=1
\documentclass[aps,prd,11pt,superscriptaddress,nofootinbib]{revtex4}

\usepackage{mathrsfs, amssymb, amsmath, amsfonts, txfonts, latexsym, graphicx}  
\usepackage[utf8]{inputenc}
	\usepackage[
      colorlinks=true,
      linkcolor=blue,
      urlcolor=cyan,
      filecolor=magenta,
      citecolor=red,
      pdfstartview=FitV,
      hyperfootnotes=false,
      pdftitle={Quantum Corner Symmetry: Representations and Gluing},
        pdfauthor={Luca Ciambelli, Jerzy Kowalski-Glikman, Ludovic Varrin},
        pdfsubject={Quantum Corner Symmetry: Representations and Gluing},
        pdfkeywords={Quantum Corner Symmetry: Representations and Gluing},
        bookmarksopen=true
      ]{hyperref}
\usepackage[all]{hypcap}
\usepackage{soul}

\usepackage{physics}

\makeatletter
\DeclareRobustCommand{\loplus}{\mathbin{\mathpalette\dog@lsemi{+}}}

\usepackage{float}
\usepackage{mathtools}

\newcommand{\dog@lsemi}[2]{\dog@semi{#1}{#2}{270,90}}
\newcommand{\dog@semi}[3]{%
  \begingroup
  \sbox\z@{$\m@th#1#2$}%
  \setlength{\unitlength}{\dimexpr\ht\z@+\dp\z@\relax}%
  \makebox[\wd\z@]{\raisebox{-\dp\z@}{%
    \begin{picture}(1,1)
    \linethickness{\variable@rule{#1}}
    \roundcap
    \put(0.5,0.5){\makebox(0,0){\raisebox{\dp\z@}{$\m@th#1#2$}}}
    \put(0.5,0.5){\arc[#3]{0.5}}
    \end{picture}%
  }}%
  \endgroup
}
\newcommand{\variable@rule}[1]{%
  \fontdimen8  
  \ifx#1\displaystyle\textfont3\else
    \ifx#1\textstyle\textfont3\else
      \ifx#1\scriptstyle\scriptfont3\else
        \scriptscriptfont3\relax
  \fi\fi\fi
}
\makeatother
\usepackage{tikz-cd}

\usetikzlibrary{arrows.meta,calc}

\usepackage{cleveref}
\usepackage{pict2e}
\usepackage{diagbox}

\newcommand{\beq}{\begin{eqnarray}}
\newcommand{\eeq}{\end{eqnarray}}
\newcommand{\beqn}{\begin{eqnarray}}
\newcommand{\eeqn}{\end{eqnarray}}

\newcommand{\RR}{\mathbb{R}}

\newcommand{\spl}[1]{\mathrm{SL}(#1,\mathbb{R})}

\usepackage{pict2e}

\newcommand{\chkM}{{\color{red} \,\checkmark\kern-5pt{}_{M}}}

\newcommand{\ee}{\end{equation}}
\newcommand{\bea}{\begin{eqnarray}}
\newcommand{\eea}{\end{eqnarray}}

\usepackage{BOONDOX-cal}

\newcommand{\da}{{a^\dag}}

\newenvironment{Align}{\begin{equation}
\begin{aligned}}
{\end{aligned}
\end{equation}\par}
\newenvironment{Align*}{\begin{equation*}
\begin{aligned}}
{\end{aligned}
\end{equation*}\par}

\usepackage{color}

\usepackage{layout}

\DeclareFontFamily{OT1}{rsfs}{}
\DeclareFontShape{OT1}{rsfs}{m}{n}{ <-7> rsfs5 <7-10> rsfs7 <10->rsfs10}{} 
\DeclareMathAlphabet{\mycal}{OT1}{rsfs}{m}{n}

\begin{document}


\title{Quantum Corner Symmetry: Representations and Gluing}

\author{Luca Ciambelli}
\affiliation{Perimeter Institute for Theoretical Physics, 31 Caroline St. N., Waterloo ON, Canada, N2L 2Y5}
\email{ciambelli.luca@gmail.com}

\author{Jerzy Kowalski-Glikman}
\affiliation{Faculty of Physics and Astronomy, University of Wroclaw, Pl. Maksa Borna 9, 50-204
Wroclaw}\email{jerzy.kowalski-glikman@uwr.pl.edu}
\affiliation{National Centre for Nuclear Research, Pasteura 7, 02-093 Warsaw, Poland}\email{ludovic.varrin@ncbj.gov.pl}

\author{Ludovic Varrin}
\affiliation{National Centre for Nuclear Research, Pasteura 7, 02-093 Warsaw, Poland}


\begin{abstract}
The corner symmetry algebra organises the physical charges induced by gravity on codimension-$2$ corners of a manifold. In this letter, we initiate a study of the quantum properties of this group using as a toy model the corner symmetry group of two-dimensional gravity $\spl{2}\ltimes R^2$. We first describe the central extensions and how the quantum corner symmetry group arises and give the Casimirs. We then make use of one particular representation to discuss the gluing of corners, achieved by identifying the maximal commuting sub-algebra. This is a concrete implementation of the gravitational constraints at the quantum level. 
\end{abstract}

\maketitle



Thanks to Noether's second theorem, gauge symmetries can acquire a non-vanishing Noether charge on codimension-2 surfaces called "corners", thereby rendering these symmetries physical. The physical symmetries and associated charges then form an algebra, known as corner symmetry algebra. One convenient way to interpret this result is the introduction of new degrees of freedom forming a representation of this algebra \cite{Donnelly:2016auv}. These fields are the \textit{edge modes} and  play an important role in quantum entanglement of spatial subregions, holography and quantum gravity \cite{Donnelly:2014fua,Donnelly:2015hxa,Speranza:2017gxd,Geiller:2017whh,Wong:2017pdm, Geiller:2019bti,Freidel:2019ees,Carrozza:2022xut,Donnelly:2020xgu,Freidel:2020xyx,Freidel:2020svx,Freidel:2020ayo,David:2022jfd,Donnelly:2022kfs,Assanioussi:2023jyq,Joung:2023doq,Mukherjee:2023ihb, Ball:2024hqe}. For gravity, further investigations revealed that diffeomorphisms realize a universal symmetry algebra on corners, independently of any (peudo)-Riemannian structure in the bulk \cite{Ciambelli:2021vnn}. This \textit{universal corner symmetry} algebra is given by 
\begin{equation}\label{universalcorneralgebra}
    \mathfrak{ucs} = \qty(\mathrm{Diff}\qty(S) \loplus \mathfrak{gl}\qty(2,\mathbb{R}))^S \loplus (\mathbb{R}^2)^S,
\end{equation}
where $S$ denotes the corner. This result is at the core of the \textit{Corner Proposal} \cite{Ciambelli:2022vot, Ciambelli:2023bmn, Freidel:2023bnj} which is an approach to (quantum) gravity centred on symmetries. In a similar fashion to the Poincaré group in quantum field theory, one expects the full representation theory of the $\mathrm{UCS}$ to provide insight into the fundamental structure of quantum gravity.\footnote{This implicitly assumes that this symmetry group  survives entirely at the quantum level, modulo potential quantum anomalies, which are allowed for these global symmetries.} This program of taking symmetries as the guiding principle is part of a bottom-up approach which aims to unlock some fundamental characteristics of quantum gravity, anchored to a controllable classical limit. While far from a full quantum gravity theory, the idea is to constrain the final theory without having to give a complete description of it from ad hoc considerations. More concretely, we expect that corner symmetries and their representations play an important role in a complete theory of quantum gravity. This is supported by their link with edge modes, which have been advocated to be necessary to correctly account for  entanglement entropy in general gauge theories, see e.g.  \cite{Anninos:2020hfj, Anninos:2021ihe, Ball:2024hqe}. For generic corners, the algebra \eqref{universalcorneralgebra} reduces to the \textit{extended corner symmetry} algebra
\begin{equation}\label{extendedcorneralgebra}
    \mathfrak{ecs} = \qty(\mathrm{Diff}\qty(S) \loplus \mathfrak{sl}\qty(2,\mathbb{R}))^S \loplus (\mathbb{R}^2)^S.
\end{equation}

One example of the application of this framework, is the proper description of localized subsystems in gravity. {This amounts to studying how corners delimitating subregions are glued together, as preliminarily discussed by Donnelly and Freidel \cite{Donnelly:2016auv}. The main challenges in describing the factorization of the Hilbert space into its subsystem components are twofold.} Firstly, the translation part $(\mathbb{R}^2)$ of the corner symmetry group moves its position. At the symplectic level they consequently generate fluxes and are thus not hamiltonian charges. The second and most important challenge comes from the infinite dimensionality of the algebra \eqref{extendedcorneralgebra}, which makes its representation theory a hard task. 

The solution to the first issue is the extended phase space formalism \cite{Ciambelli:2021nmv, Freidel:2021dxw}, where edge modes are understood as a dynamical embedding map which resolves the flux problems and makes the translation charges integrable. We therefore use the full $\mathfrak{ecs}$ in our analysis. In fact, one of our major results is that that  translations are needed for consistency of the gluing procedure. 

In order to tackle the second issue, we will focus on the   finite part of the algebra \eqref{extendedcorneralgebra}, which can be viewed both as a toy model and one of intrinsic interest. The finite part of the ECS is the five dimensional semi-direct product group
\begin{equation}\label{ecsgroup}
    \mathrm{ECS} = \spl{2} \ltimes \mathbb{R}^2.
\end{equation}
This group is the extended corner symmetry group of $2D$ gravity, where the corner is simply a point. Moreover, this group is realized also in higher dimensions per point on the corner. Therefore, the study of its quantum representations is at the core of the corner proposal, and these representations will inform more sophisticated analysis in the future. {Reintroducing diffeomorphisms leads to an infinitely many copies of this algebra, one per point on the corner. Diffeomorphisms act non-trivially on $\spl{2} \ltimes \mathbb{R}^2$, enforcing covariance of local expressions. Upon suitably integrating on the corner, we  expect the Casimir analysis performed here to pertain to the more general case.} {In the present work, we give the first concrete application of the corner proposal: Using the representation theory of the finite part of the $\mathrm{ECS}$, we propose a novel explicit Hilbert space realization of the gluing procedure of corners and their algebras}.

The quantum states live in projective representations of the symmetry group, which in turn are equivalent to the representations of the maximally centrally extended group \cite{Bargmann_1954,Mackey_1958}. As a result, in the quantum case, one finds a central extension of the $\mathrm{ECS}$, and  we call this new algebra the \textit{quantum corner symmetry} algebra ($\mathrm{QCS}$). Together with the gluing procedure, this constitutes the main message of this letter. Together, these results suggest that the corner proposal can be rendered rigorous at the quantum level, thus providing a viable bottom-up approach to quantum gravity.

The algebra associated to the group \eqref{ecsgroup} is generated by five operators $L_0,L_\pm,P_\pm$. The subset $L_0,L_\pm$ generates the special linear algebra
\begin{equation}
     [L_0,L_\pm] = \pm L_\pm, \qquad [L_-,L_+] = 2 L_0,
\end{equation}
while the cross commutation with the translation part is given by
\begin{equation}
     [L_0, P_\pm] = \pm \frac12 \, P_\pm\,,\quad [L_\pm, P_\pm] =0\,,\quad [L_\pm, P_\mp] =\mp P_\pm.
\end{equation}
The algebra admits one cubic Casimir \cite{Ciambelli:2022cfr}
\begin{equation}\label{ecscasimir}
    \mathcal{C}^{(3)}_{\mathrm{ECS}} =  L_+ P_-^2 + L_- P_+^2 -2 L_0 P_- P_+.
\end{equation}

The $\mathfrak{ecs}$ algebra possesses a (unique) non-trivial co-cycle in the translation sector. We thus extend the algebra by the central term
\begin{equation}\label{centralextension}
    \qty[P_-,P_+] = C\,,
\end{equation} 
with $C$ having dimension of inverse length square.
The Abelian algebra of normal translations is thus replaced by the Heisenberg algebra. 
 We call the centrally-extended corner symmetry algebra obtained \textit{Quantum Corner Symmetry} (QCS) algebra 
\begin{equation}\label{QCA}
    \mathfrak{qcs} =\mathfrak{sl}(2,\mathbb{R}) \loplus \mathfrak{h}_3,
\end{equation}
where $\mathfrak{h}_3$ denotes the three dimensional Heisenberg algebra generated by $P_\pm$ and $C$. 

The $\mathfrak{ecs}$ algebra can be derived from a gravitational phase space. The charge algebra is a projective representation of the symmetry algebra. Consequently, the central extension in the translations sector has been found in various previous works, such as \cite{Afshar:2016wfy, Speranza:2017gxd, Chandrasekaran:2020wwn}. We conclude that the $\mathrm{QCS}$ algebra should be taken as the starting point in the  study of quantum representations of the corner algebras. 

To do so, it is of utmost importance to classify the Casimir operators. Since they commute with all other generators, Schur's lemma implies that, in an unitary irreducible representation, the Casimirs act as a multiple of the identity. The multiplication factor depends on the weights of the representation and will thus be different for non-equivalent representations. As such, this scalar value of the Casimirs labels the different unitary irreducible representations. The quantum field theory framework gives the most famous example of such phenomenon in Wigner's classification of the Poincaré group's unitary irreducible representations \cite{Wigner:1939cj}. 

It is thus evident that the first step in understanding the $\mathrm{QCS}$ is to describe its Casimirs. The central element introduced above is trivially the first one, as it commutes with all other generators. The second one is given by a modification of the cubic Casimir of the non-centrally extended algebra \eqref{ecscasimir}
\begin{align}\label{qcscubiccasimir}
    \mathcal{C}^{(3)}_{\mathrm{QCS}} &= C\qty(L_0(2 L_0 + 3)-2 L_- L_+) +L_- P_+^2 + L_+ P_-^2 -2 L_0 P_- P_+.
\end{align}
One can show that these are the  only Casimirs of the $\mathrm{QCS}$, see section 4 of \cite{Varrin:2024sxe}.

We see that the cubic Casimir of the $\mathrm{ECS}$ is recovered in the limit where the central element vanishes
\begin{equation}
    \lim_{C \mapsto 0} \mathcal{C}^{(3)}_{\mathrm{QCS}} = \mathcal{C}_{\mathrm{ECS}}^{(3)}.
\end{equation}
In \cite{Ciambelli:2022cfr}, this Casimir is shown to be activated in the $\mathfrak{ucs}$ by the $GL(2,\RR)$ singlet, which is generated by conformal rescalings of the corner. This fact suggests that this Casimir encodes the value of the conformal energy at the corner, which remains constant on the orbit action of the $\mathfrak{qcs}$.


Let us now turn to the representation theory of the $\mathrm{QCS}$. While the individual unitary irreducible representations of both the $\spl{2}$ part and the Heisenberg part are known, the representation of the semi-direct product is far from trivial. The method of inducing known representations of a subgroup to the entire group is known as Mackey theory \cite{Mackey1951,Mackey1952,Mackey1953}. More specifically, in the case of the semi-direct product like the Poincaré group, Wigner and Bargmann developed a method to find all the unitary irreducible representations using the so-called little group analysis \cite{Wigner:1939cj,Wigner1948,Wignerbargman1948}.

However, the construction of the irreducible representations in the case of a non Abelian normal subgroup is far from trivial. While this letter focuses on gluing corners, we will address this topic in a separate work \cite{Varrin:2024sxe}. For now, we therefore induce the Fock representation of the Heisenberg group to the rest of the $\mathrm{QCS}$. In doing so, we obtain the metaplectic representation of the special linear group which, when acting on the Heisenberg group, becomes irreducible. The Stone-von Neumann theorem \cite{vN1, Stone1}, states that the Fock representation is unique. This consequently assures that the corresponding $\mathrm{QCS}$ representation is the unique one acting on this Hilbert space. However, one cannot disregard the possibility of the existence of other unitary irreducible representations on a different Hilbert space that does not reduce to a unitary irreducible representation when restricted to the Heisenberg part.\footnote{We thank Jos\'e Figueroa-O'Farrill for correspondence on this point.}

Consider thus the unitary irreducible representations of the Heisenberg algebra $\mathfrak{h}_3$. We begin by requiring that the central element acts as a multiple of the identity on the Hilbert space of the representation\footnote{The constant is chosen to be real in order to ensure the unitarity of the corresponding group representation.}
\begin{equation}
    C \ket{\psi} = c \ket{\psi}, \quad c\in \mathbb{R}\setminus \qty{0}.
\end{equation}
Stone-von Neumann's theorem  states that there exists a unique unitary irreducible representation of $\mathfrak{h}_3$ for each $c$. The corresponding Hilbert space can be described by the Fock space of the quantum harmonic oscillator
\begin{equation}
    \mathcal{H} = \qty{\ket{n},n\in \mathbb{N}},
\end{equation}
with the usual scalar product
\begin{equation}\label{scalarproduct}
    \braket{n}{m} = \delta_{nm}.
\end{equation}
The Heisenberg algebra can be realized in this representation as{\footnote{From this point we focus on the $c > 0$ case. Note that choosing $c<0$ instead would invert the roles of $P_\pm$ as the creation/annihilation operators.}
\begin{equation}\label{metaplecticR2}
    P_- = \sqrt{c} a,\quad
    P_+ = \sqrt{c} \da,\quad
    C = c,
\end{equation}
where $a$ and $\da$ are the usual creation and annihilation operators satisfying $[a,\da]=1$. 
One shows that applying such operators to states preserves the positive-definiteness of the norm induced by eq.~\eqref{scalarproduct}. Thus, the group representation obtained by exponentiation is  unitary.

Inducing this representation to the special linear part we obtain the metaplectic representation
\beq\label{metaplecticsl2r}
L_+=\frac12 \da\da,\quad L_-=\frac12 a a,\quad L_0=\frac12 \da a+\frac14.
\eeq
Note that the metaplectic representation of $\mathfrak{sl}(2,\RR)$ defined by equation \eqref{metaplecticsl2r} is not, by itself, irreducible. However, the inclusion of the Heisenberg group connects the odd and even numbered states and makes it into an irreducible representation of the $\mathrm{QCS}${, in which $\mathcal{C}^{(3)}_{\mathrm{QCS}}=\frac38 c$}. It is in this regard that translations play a prominent role in our analysis.

One can use the explicit expressions of the generators in this representation to calculate the cubic Casimir \eqref{qcscubiccasimir}. As expected from Schur's lemma and the irreducibility of the representation, its action on the Fock space is equivalent to simply multiplying by a scalar. We therefore obtained a family of irreducible unitary representations of the $\mathrm{QCS}$ labeled by $c \in \mathbb{R}\backslash\qty{0}$. Note that taking the limit $c\mapsto 0$ does not produce a representation of the $\mathrm{ECS}$. In fact, it can be proven that it is impossible to reduce the oscillator representation of the special linear algebra \eqref{metaplecticsl2r} to the Abelian translations of the $\mathfrak{ecs}$: this representation exists solely because of the quantum non-commuting nature of the translation sector. The unitary irreducible representations of the $\mathrm{ECS}$ are therefore of a different nature than the oscillator representation used here \cite{Varrin:2024sxe}. 

The physics of this Hilbert space of  $2$-dimensional gravity is as follows. Inspired by the covariant phase space analysis, {the  $\mathrm{QCS}$ group is the organizing principle of corner charges, at the quantum level}. In gravity, the latter are a collection of metric components at and in the vicinity of the corner. Therefore, the Hilbert space we constructed is the quantum version of the geometric data at a corner. This is where this bottom-up approach typical of the corner proposal is far-reaching: without having to describe a complete quantum gravity theory, we immediately see the need for the appearance of a quantum geometry description of the corner. Then, the $\mathrm{QCS}$ operators act on this Hilbert space creating or annihilating "quantum bits of geometry". In this sense, the vacuum state should be thought of as absence of geometry, rather than a flat background. Another advantage lies in the fact that only the {corner algebra is promoted to an organizing principle of quantum operators}. This is the power of being anchored to symmetries: the metric itself is not a fundamental quantum datum, only the charge algebra is. While this is a preliminary and informal discussion, we expect that exploring the quantum nature of gravity in this framework can be much rewarding, and we are here initiating this avenue of research.
\newline

We now turn to the application of the representation theory developed in the previous sections to the gluing of two corners, which can be regarded as the endpoint of two subsystems. To define  a localized quantum subsystem in gauge theories, the initial data on the Cauchy slice has to obey gauge constraints and the data on the interior and exterior slice cannot be specified independently. This results in the failure of the tensor product factorization of the associated Hilbert spaces. {Indeed, the entangling surface --the shared boundary-- is a corner and as such it carries a representation of the corner algebra. This has been described by Donnelly and Freidel in \cite{Donnelly:2016auv}, where the authors introduced new degrees of freedom living on the boundary,  the edge modes, carrying a representation of the corner symmetry group. In the quantum case, the representation carried by the edge modes is used to define the gluing procedure between the Hilbert spaces of the individual subregions. Here, we are interested in studying how the representations of the corner algebra for the two corners are glued together into a single algebra.}

In the case at hand, the representation theory of the corner symmetry group is taken to describes the quantum states of the edge modes. The gluing procedure can thus be explicitly constructed as a condition on the tensor product of the states described in the previous sections.
Consider a spacelike segment $\Sigma_L$ connected to the corner $L$ as depicted in Figure \ref{fig:onesegment}.
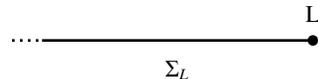
\begin{figure}[H]
    \centering
 \begin{tikzpicture}
  \draw[line width=1pt, dotted] (0,0) -- (.4,0); 
  \draw[line width=1pt] (.4,0) -- (4,0) node[midway, below,yshift = -4pt] {$\Sigma_L$}; 
  \fill (4,0) circle (2pt) node[above,yshift = 4pt]{L}; 
\end{tikzpicture}
    \caption{One spacelike segment connected to the boundary $L$. The Hilbert space describing this system is a representation of the corner symmetry group at $L$.}
    \label{fig:onesegment}
\end{figure}
To the corner of this Cauchy surface, we associate a Hilbert space $\mathcal{H}_L$ carrying a unitary irreducible representation of the corner symmetry group. Let us now consider a second segment $\Sigma_R$ with corner $R$ and its own Hilbert space $\mathcal{H}_R$. Identifying the two corners and gluing the two segments together, as depicted in Figure \ref{fig:gluing}, yields a Hilbert space $\mathcal{H}_G$ associated with a point of the glued segment. Note that the presence of a Hilbert space associated with a bulk point is necessary, since any point can be taken as a fictitious boundary from which one can split into two subregions. The glued Hilbert space is not the tensor product of the left and right ones, but a proper subspace
\begin{equation}\label{embeddinggluedhilbertspace}
    \mathcal{H}_G = \mathcal{H}_L \otimes_{\mathrm{QCS}} \mathcal{H}_R \subset \mathcal{H}_L \otimes \mathcal{H}_R = \tilde{\mathcal{H}}_G,
\end{equation}
where $\otimes_{\mathrm{QCS}}$ denotes the entangling product defined by the corner symmetry group $\mathrm{QCS}$, whose construction is given in what follows.

How do we implement this gluing procedure at the quantum level? We here propose that the quantum version of the gravitational constraints is the requirement that the maximal commuting sub-algebra of the two $\mathrm{QCS}$ should match. Eventually, we must obtain that the glued segment holds a representation of the corner symmetry group. In order for the whole procedure to be consistent, the glued representation should be constructed from the available left and right operators.

Let us elucidate the process of segment gluing through an analogy with the construction of Feynman diagrams. Consider a three-valent vertex in a Feynman diagram, where the incoming momentum is  $p$ and the outgoing momenta are $q$ and $r$. Momentum conservation at the vertex is ensured by the inclusion of the delta function $\delta(p-q-r)$. 
Now, let us turn to a bi-valent vertex, involving one incoming and one outgoing particle with momenta $p$ and $q$, respectively. Although such a vertex may appear trivial, it must be incorporated for consistency. The delta function associated with this vertex is $\delta(p-q)$, reflecting the equality of the incoming and outgoing momenta.
The momenta in this context are the eigenvalues of momentum operators, which are tied to the Noether charges corresponding to translational symmetry at the endpoints (or corners) of the lines. A natural question arises: why do we impose momentum conservation at the vertex, rather than equating other charges? The answer lies in the structure of the Poincaré algebra: momentum operators constitute its maximal commuting subalgebra. 
To summarize, in the case of a bi-valent vertex in Feynman diagram construction, we impose the equality of the eigenvalues of the maximally commuting set of operators within the Poincaré algebra. Our approach in the context of corner symmetry reflects this same physical intuition, adapted to the relevant framework.

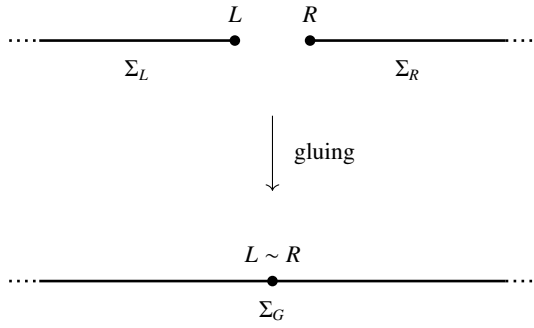
\begin{figure}
    \centering
    \begin{tikzpicture}
        \draw[line width=1pt, dotted] (0,0) -- (.4,0); 
        \draw[line width=1pt] (.4,0) -- (3,0) node[midway, below,yshift = -4pt] {$\Sigma_L$}; 
        \fill (3,0) circle (2pt) node[above,yshift = 4pt]{$L$}; 
        \draw[line width=1pt] (4,0) -- (6.6,0) node[midway, below,yshift = -4pt] {$\Sigma_R$};
        \draw[line width=1pt,dotted] (6.6,0) -- (7,0);
        \fill (4,0) circle (2pt)  node[above,yshift = 4pt]{$R$};

        \draw[->] (3.5,-1) -- (3.5,-2) node[midway,right]{\hspace{5pt}\small{gluing}};

        \draw[line width=1pt, dotted] (0,-3.2) -- (.4,-3.2);
        \draw[line width=1pt] (.4,-3.2) -- (6.6,-3.2) node[midway, below,yshift = -4pt] {$\Sigma_G$};
        \draw[line width=1pt,dotted] (6.6,-3.2) -- (7,-3.2);
        \fill (3.5,-3.2) circle (2pt) node[above,yshift = 4pt]{\small $L \sim R$};
        
    \end{tikzpicture}
    \caption{Two segments are glued together identifying the left and right corners. The Hilbert space associated with the Cauchy slice $\Sigma_G = \Sigma_L \cup \Sigma_R$ is given by a proper subspace of the tensor product of the left and right Hilbert space.}
    \label{fig:gluing}
\end{figure}

As an aside, one can show that the translation sector, corresponding to the Heisenberg subgroup of the $\mathrm{QCS}$, is \textit{necessary} to describe the factorization of sub-regions. This is due to the fact that with the $\spl{2}$ algebra we only have access to operators quadratic in the oscillators, and therefore the procedure is not self-consistent without translations. Classically, this can be understood as the fact that translations physically move the two corners, a necessary condition to glue them together. Of course, this problem does not arise when considering a single corner, but it is pertinent to gluing.

 
In order to perform the gluing, we need to identify the maximal commuting sub-algebra of the $\mathrm{QCS}$, which is $3$ dimensional. We can perform a change of basis at the algebra level which facilitates the physical interpretation of the gluing procedure. Let us define the following self-adjoint operators
 \begin{Align}
      X &= \frac{C^{-1}}{\sqrt{2}} \qty(P_+ + P_-),\quad P =  \frac{i}{\sqrt{2}}(P_+ - P_-),\\
     V &= L_0 +\frac12\qty( L_+ + L_-),\quad K =  L_0 - \frac12(L_+ + L_-),\\ 
     L &= L_+-L_-.
 \end{Align}
\noindent Once we write them in the metaplectic representation (\ref{metaplecticR2},\ref{metaplecticsl2r}), the notation becomes clear: The $X$ and $P$ operators correspond to the position and the momentum of the harmonic oscillator whereas $K$ and $V$ correspond respectively to the kinetic and potential terms.  Curiously, these three operators correspond to the three radial conformal Killing vectors on Minkowski space: $K$ is the Hamiltonian, V is the conformal boost, while $L$ is the dilation \cite{Arzano:2021cjm}. This might be more than a mere coincidence, and deserves further exploration. Further, we interpret  $X$ and $P$ as defining at the quantum level the notion of position and momentum of the corner. This interpretation is also consistent with the physics of edge modes, which are indeed the embedding fields defining the corner's position.

One can choose the maximal commuting sub-algebra to be either $(C,X,V)$ or $(C,P,K)$. We will focus on the former. Consider a left and a right copy of the Fock space and their algebras. In the position eigenstates basis we have
\begin{equation}
    \ket{x}_{L/R} = \sum_{n=0} \psi_n(x) \ket{n}_{L/R},
\end{equation}
where $\psi_n(x)$ is the wave function of the harmonic oscillator expressed through Hermite polynomials, that by construction are a function of $\sqrt{c} x$, which is dimensionless. Then we have
\begin{align}
    C^{L/R} \ket{x}_{L/R} &= c_{L/R} \ket{x}_{L/R},\label{CLR}\\
    X^{L/R} \ket{x}_{L/R} &= x_{L/R} \ket{x}_{L/R},\label{XLR}\\
    V^{L/R} \ket{x}_{L/R} &= c_{L/R} x_{L/R}^2 \ket{x}_{L/R}.\label{VLR}
\end{align}

In order to describe the gluing, we start from the pre-Hilbert space
\begin{equation}
    \tilde{\mathcal{H}}_G = \qty{\ket{x}_L\otimes \ket{y}_R}.
\end{equation}
The glued Hilbert space is then obtained by equating the left and right action of the maximal commuting sub-algebra. Looking at \eqref{CLR}, the central element must be equated, forcing the left and right representations to be the same. Furthermore, eqs. (\ref{XLR},\ref{VLR}) impose the glued states to be diagonal in the position basis,
\begin{equation}
    \mathcal{H}_G = \qty{\ket{x}_G = \ket{x}_L \otimes \ket{x}_R}.
\end{equation}
The gluing condition on corners that have a definite position is simply that they are at the same position. We could have instead used the maximal commuting sub-algebra $C,P$ and $K$. Then, the glued Hilbert space would have been diagonal in the momentum basis. The equivalence of the two choices is due to the equivalence of the position and momentum basis. Independently of the chosen basis, the gluing is a continuity condition for the variables describing the system.

We can further construct the glued $\mathrm{QCS}$ algebra on the glued Hilbert space. The glued operators are functions of the left and right operators given explicitly by
\begin{Align}\label{glued algebra}
    X_G &= \frac{1}{\sqrt{2}}\qty(X_L + X_R),\quad  P_G = \frac{1}{\sqrt{2}}(P_L + P_R),\\
    V_G &=  \frac12\qty(V_L + V_R + C X_L X_R), \quad K_G = \frac12\qty(K_L + K_R + \frac{P_L P_R}{C}),\\
    L_G & = \frac12\qty(L_L + L_R - i  \qty(X_R P_L + X_L P_R)),
\end{Align} 
\noindent where we have used that the left and right central elements are equal and called $C$. Although the irreducibility of this representation can be checked explicitly, it follows simply from the fact that the special linear operators can be expressed as quadratic expressions of the translations, similar to the left and right cases. In fact, the glued cubic Casimir acting on the glued Hilbert space gives the same constant value as the left and right ones. This confirms that the glued representation is indeed the same as the ones we started with.

Once the gluing understood, the split of one corner into two follows the exact opposite path. Start with a spatial segment and choose the corner point at which split the system. There is now one copy of the Hilbert space associated with the chosen corner. The system is doubled taking the diagonal tensor product and then relaxing the gauge constraints,
\begin{equation}\label{splittingprocedure}
    \ket{x} \xlongrightarrow{\text{double}} \ket{x}_L \otimes \ket{x}_R \xlongrightarrow{\text{relax}} \ket{x}_L \otimes \ket{y}_R.
\end{equation}
Thus, starting from a continuous Cauchy slice without boundaries we have a way to separate it into two independent localized subsystems.


Let us recap our findings and conclude. We have found the unique central extension of the $\mathfrak{ecs}$, and argued for its quantum origin. We then constructed the unique unitary irreducible representation induced from the Stone-von Neumann representation of the Heisenberg algebra. Using this representation we described the gluing procedure of two corners by explicitly constructing the entangling product. We thus laid the ground work for the study of entanglement entropy between two spatial subregions in quantum gravity within the framework of the corner proposal. While a comprehensive study of this topic will be the focus of upcoming works, we remark that $L_0$ is ready-made to be interpreted as the modular Hamiltonian for the subregion algebra \cite{Arzano:2021cjm}, with $c$ setting the scale of the problem, and thus related to temperature. The natural appearance of a scale from the central extension of the corner algebra is a remarkable feature, whose consequences are yet to be unveiled.

There are several future directions. One of them is to study the coadjoint orbit method of Kirillov \cite{Kirillov_1962,Kirillov1976, Kirillov1990,Kirillov1999,Kirillov2004}, that could offer insights into the description of the representations. This would echo and complement \cite{Donnelly:2020xgu, Ciambelli:2022cfr, Donnelly:2022kfs}.

Another direction is to restore the corner diffeomorphisms. In particular, one should study the fate of the central extension and the cubic Casimir in order to understand if there is a consistent way to promote the analysis of this work to each point on the corner in higher dimensions. The simplest generalization of this setup occurs in $3D$, where the corner is a topologically a circle. The study of central extensions in this case is ongoing, and we anticipate that the Virasoro algebra plays a crucial role. More generally, in higher dimensions, we expect that the area-preserving diffeomorphisms of the corner are central to understanding both the structure of central extensions and their quantum representations, see the analysis in \cite{Donnelly:2020xgu, Donnelly:2022kfs}.

Additionally, we wish to connect with top-down $2D$ quantum gravity models. The generalization of the gluing procedure to a higher number of corners is straightforward, and is reminiscent of the setup of Causal Dynamical Triangulation, see 
\cite{Ambjorn:1998xu, Loll:2019rdj}.

Finally, there are some differences between the entangling product proposed here and the one used in \cite{Donnelly:2016auv}. Indeed, our construction relies on the analysis performed in \cite{Ciambelli:2021nmv}. We plan to report on the relationship between these entangling products in a future publication.

\vspace{1cm}

\textit{Acknowledgements:} We are grateful to Michele Arzano, Adam Ball, Jackie Caminiti, Federico Capeccia, Laurent Freidel, and Rob Leigh for useful discussions. JKG and LV thank Perimeter Institute for hospitality. Research at Perimeter Institute is supported in part by the Government of Canada through the Department of Innovation, Science and Economic Development Canada and by the Province of Ontario through the Ministry of Colleges and Universities. For JKG, this work was supported by funds provided by the National Science Center, project number 2019/33/B/ST2/00050.

\bibliographystyle{uiuchept}
\bibliography{file.bib}

\end{document}